# GPS Time Synchronization in School-Network Cosmic Ray Detectors

Hans-Gerd Berns, Toby H. Burnett, Richard Gran, R. Jeffrey Wilkes

?

*Abstract*—The QuarkNet DAQ card for school-network cosmic ray detectors provides a low-cost alternative to using standard particle and nuclear physics fast pulse electronics modules. The board, which can be produced at a cost of less than US$500, produces trigger time and pulse edge time data for 2 to 4-fold coincidence levels via a universal RS232 serial port interface, usable with any PC.

Individual detector stations, each consisting of 4 scintillation counter modules, front-end electronics, and a GPS receiver, produce a stream of data in form of ASCII text strings in identifiable set of formats for different functions. The card includes a low-cost GPS receiver module, which permits time-stamping event triggers to about 50 nanosecond accuracy in UTC between widely separated sites. The technique used for obtaining precise GPS time employs the 1PPS signal, which is not normally available to users of the commercial GPS module. We had the stock model slightly custom-modified to access this signal. The method for deriving time values was adapted from methods developed for the K2K long-baseline neutrino experiment. Performance of the low-cost GPS module used is compared to that of a more expensive unit with known quality.

## I. INTRODUCTION

SEVERAL projects around the world are aiming at building large-scale ultra-high-energy cosmic ray detector networks using secondary schools as detector sites. Three groups began a collaboration in 2001 to design and build a low-cost data acquisition (DAQ) module [2] for distribution to high school networks: QuarkNet [3], based at Fermilab; the Washington Area Large-scale Time-coincidence Array (WALTA) [4][5], based at the University of Washington, Seattle; and the Cosmic Ray Observatory Project (CROP) [6], based at the University of Nebraska, Lincoln. One of the requirements for the DAQ module design was to implement accurate time-stamping of each detector trigger, with accuracy on the order of 1 microsecond or better in absolute Universal Time (UTC), as a means to synchronize the cosmic ray data at widely separated school sites. The Global Positioning System (GPS) [7] provides an ideal method of synchronized tracking around the world at a low cost for the user. This paper will describe how the GPS time synchronization is implemented in the QuarkNet DAQ system. Details about the DAQ card are also available in these proceedings [2].

## II. GPS TIMING

The Global Positioning System consists of at least 24 satellites maintained by the US Department of Defense (DOD), each transmitting coordinated "GPS Time" according to its onboard atomic clock. GPS Time differs from UTC only in the absence of the leap seconds, which are periodically inserted in UTC. Most GPS receivers (including ours) automatically take the shift into account using data downloaded from the satellites, so time reported is UTC. The satellites' onboard clocks are regularly conditioned to match GPS time according to a ground-based reference clock system (actually a large number of high-precision atomic time standards). The satellites also broadcast their ephemerides, so their position in space can be accurately calculated as a function of time. The ephemerides also are regularly recalculated and updated. With each satellite's position in space known to high accuracy from its ephemeris, users' receivers can fit for their position and time (x,y,z,t) if four or more satellites are simultaneously in view.

Since the GPS satellites are constantly referenced to a national standards laboratory time base, the GPS system provides a simple and inexpensive way to obtain high precision absolute time, synchronized to UTC, without purchasing and constantly recalibrating a set of atomic clocks. The GPS system is designed to give standard errors of about 150 meters on a single position fit and 150 nanoseconds relative to UTC on a single time fit. For a fixed antenna location, long-term averaging of the measured antenna coordinates provides improved accuracy.

## III. GPS RECEIVER: LEADTEK GPS-9532W

We adapted a low-cost commercial GPS receiver, Leadtek Research, Inc. model GPS-9532 [8]. The basic specifications are as follows:

- ?? 12-channel GPS receiver and antenna in one unit.
- ?? Small, light weigh and weatherproof package: approx. 63 mm diameter, 20 mm height, and 3 meter cable.
- ?? Special modifications for the QuarkNet DAQ (W-suffix):

Manuscript received November 14, 2003. This work was supported in part by the U.S. Department of Energy and QuarkNet.

The Authors are with the University of Washington, Physics Department, Seattle, WA 98195-1560, USA.
  H. G. Berns   (tel.: 206-685-4725, e-mail: berns@phys.washington.edu).
  T. H. Burnett (tel.: 206-543-8963, e-mail: tburnett@u.washington.edu).
  R. W. Gran    (tel.: 206-543-9584, e-mail: gran@phys.washington.edu).
  R. J. Wilkes  (tel.: 206-543-4232, e-mail: wilkes@phys.washington.edu).

- 1-pulse-per-second (1PPS) signal available at output and enabled, i.e. power savings disabled at power-up.
- Single DB-9 connector for all signals: serial data, 1PPS, and 5VDC supply power
- Default NMEA serial output at 9600 baud.

?? Low cost: approx. US$100 each (at qty. 250).

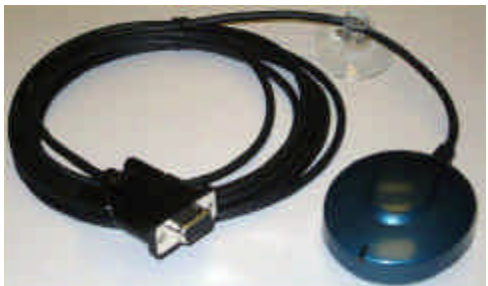

Fig. 1. GPS receiver and antenna module, in a weatherproof package, and a 3-meter cable providing supply power and data connection.

Our customized units had the 1PPS signal made available. We prepared a custom mating connector to deliver the 1PPS signal over 30-meter cables, and power the GPS units from the DAQ card instead of a separate external power supply.

## IV. EVALUATION TEST RESULTS

A sample Leadtek GPS-9543 module was tested for 1PPS jitter and accuracy at the WALTA test stand over a ~52-hour period in December 2001. The test stand was set up with a Motorola UT-Plus Oncore [9] as a reference GPS module. An equivalent Motorola Oncore receiver has already been used successfully as a backup precision timing source for the Super-Kamiokande and K2K project for several years, and this unit is known to be accurate within approx. 70 nanoseconds rms (+/- 400 ns worst case) of absolute UTC [1]. The 1PPS pulse arrival times of the two GPS receivers were simultaneously recorded via a 50-MHz VME clock counter module. The antennae of both receivers were placed close to each other at a location on the University of Washington campus where the full sky view essentially clear, perfect for GPS satellite tracking.

The evaluation test revealed some minor bugs and unanticipated "features" of the customized device:

*A. Always in Navigation Mode:* The low-cost Leadtek GPS receiver has no option to run in "position-hold" mode, but always runs in "navigation" mode. This means the unit continuously fits its position, and so requires a minimum of 4 GPS satellites tracked simultaneously at all times to continuously recalculate the position and timing solution. This tracking condition is often not satisfied, and performance is known to be sensitive to electrical noise (e.g. radio or TV towers), other environmental conditions, and even reflections from high-rise buildings nearby. At a stationary location with known 3-D position parameters, a GPS receiver in "position-hold" mode would only need to compute the timing solution, and thus only require a single GPS satellite tracked at a time

*B. 1PPS and serial data unsynchronized:* Also the 1PPS pulse and serial data stream of the Leadtek unit turn out to be unsynchronized, and in addition, the 1Hz period of the serial data output was measured to be slightly higher than 1Hz (also confirmed with all other units purchased later), i.e. drifting away from the 1PPS pulse edge time by 1 millisecond every 20-23 seconds (or a full-period shift every 5.5~6 hours). However, for our application, since the serial data are used only to determine the time down to integer seconds, this problem can be compensated in user software.

*C. 1PPS glitches:* The GPS module showed periodic 100 millisecond offset jumps every ~32 minutes, and skipped a 1PPS pulse every ~5.3 hours (every 10th occurrence of the ~32-minute glitch). Later production units, showed similar 1PPS irregularities with slightly varying periodicies between 30-35 minutes. This is probably related to the 1PPS-vs-1Hz asynchronous shift problem mentioned above. Therefore, additional error detection and correction methods have been implemented in our DAQ firmware and offline software.

On the other hand, the Leadtek device showed superior short-term 1PPS stability than the Motorola device, i.e. less jitter, and faster reacquisition times after power interruptions.

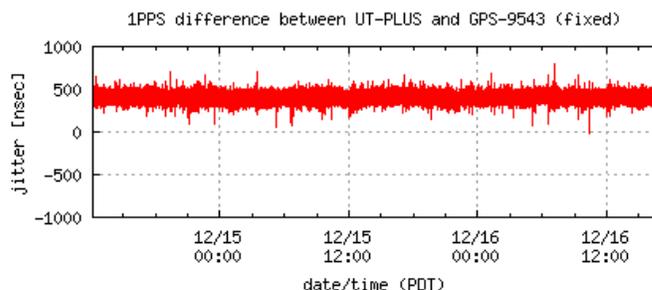

Fig. 2. Time series of the difference between Leadtek GPS-9543 and Motorola UT-Plus Oncore 1PPS arrival times, in a 52-hour test period.

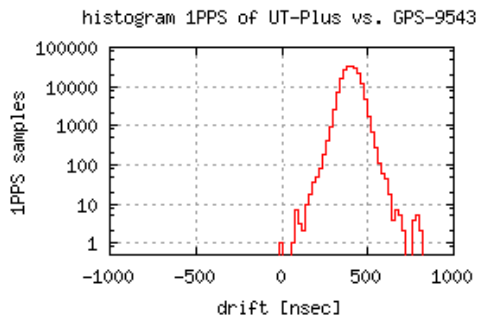

Fig. 3. Histogram of time series shown in fig.2

Comparison result: Figs. 2 and 3 show that the 1PPS of both the Motorola UT-Plus Oncore and the Leadtek GPS unit agree very nicely after the drift and jump problems mentioned above are corrected in software. The ~390 nanosecond offset in the plots comes from internal propagation delays in the receivers which were not compensated during the test period and can be ignored here. The standard deviation is approx. 45 ns, and 95.6% of the pulses are within ±100 ns of the mean.

## V. IMPLEMENTATION IN THE QUARKNET DAQ CARD

The Leadtek GPS-9532W module is connected to the QuarkNet DAQ card via a custom cable, specially made for the DAQ card, with a male 10-pin RJ-45 connector at the DAQ card end and a female 9-pin D-sub (DB9) connector at the GPS receiver end. Inside the DB9 connector housing is a small circuit board with RC filters for the 5VDC supply voltage for the GPS receiver and RS232 serial signals, and an LVDS line driver for the 1PPS pulse, to allow transmission of the signals over long cable lengths up to 100 feet without degradation. In addition, the circuit board contains a temperature sensor to allow monitoring of the outdoor temperature at the GPS receiver.

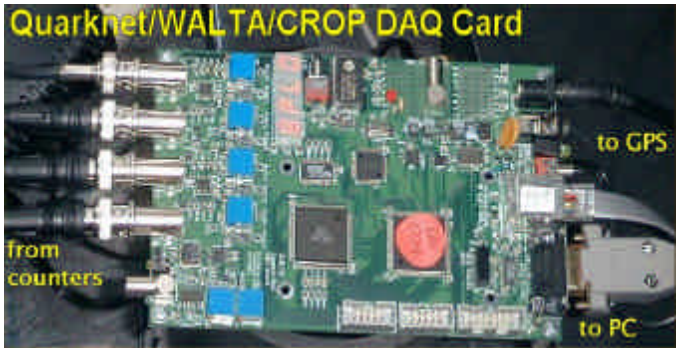

Fig. 4. QuarkNet DAQ card and cabling [2].

After power-up, the GPS receiver immediately scans the sky for suitable GPS satellite signals and by default goes into "navigation" mode, automatically self-surveying its own position (latitude, longitude and altitude) and time. No user action is needed to set the survey and power modes. Depending on the satellite conditions and previous operating conditions, the receiver might need several minutes to lock onto its position and time accurately. As soon as it tracks the first satellite, the 1PPS signal becomes active. This will initialize the CPLD counter on the DAQ card, which is used for logging the time of event triggers with 24-nanosecond accuracy [2]. However, the timing will not be accurate until it tracks at least 4 satellites, which is the minimum for calculating all parameters for 3-D positioning and UTC offset.

The user can check the GPS reception status via the DAQ card 'DG' command (Display GPS status). The 1PPS timing is unreliable if the GPS status shows an 'invalid' flag ("V" flag in the NMEA data stream). Once the GPS flag becomes 'valid' ("A" NMEA flag), the 1PPS signal will usually stay accurately aligned to absolute UTC with an estimated error of 50 nanoseconds rms or better, and DAQ operations may begin.

## VI. EVENT TIME-STAMPING

Two types of data are periodically transmitted from the GPS receiver to the DAQ card: a 1PPS square-wave signal whose leading edge is synchronized with the beginning of UTC seconds to within ~50 ns rms typically, and a serial ASCII data stream (NMEA format at 9600 baud by default) containing complete GPS data such as latitude, longitude, altitude, date, time down to milliseconds, and housekeeping data.

On the DAQ card, each 1PPS signal leading edge is latched by clock counter logic in the CPLD and used to calibrate the counter frequency. The CPLD is clocked at 41.667 MHz, generated by a 12.50 MHz temperature-compensated quartz oscillator and a 1:3.333 PLL frequency multiplier in series. Thus the timing resolution is 24 nanoseconds, and with 32-bit register width, the CPLD clock counter rolls over every 103 seconds.

The GPS serial data is processed by the microcontroller (MSP430) on the DAQ card through its secondary UART port. As mentioned above, the serial NMEA stream of the GPS-9532W device is not sent synchronously with the 1PPS pulse stream, but is updated at an internal frequency slightly higher than 1 Hz. As a result, the start of the serial data stream and its time information keeps drifting away from the 1PPS pulse at a slow rate, approx. 1 ms shift every 20-23 seconds, which makes alignment of these two data streams difficult for the user. Therefore, additional code had to be implemented in the DAQ firmware to measure the offset drift, so the serial data can correctly be aligned with the most recent 1PPS pulse and vice versa.

At each event trigger, the DAQ card records the current CPLD clock counter value (trigger time). These data are immediately sent out the primary serial port in hex format, appended with 1PPS and GPS data, at least one string per trigger, with additional strings as needed for multiple pulse edges or command responses.

```
C8B8E2A0 A2 3B 01 01 00 01 00 01 C8033BA6 212554.156 121003 A 08 0 -0266
CCDCB602 B6 01 00 01 00 01 00 01 CA7F0414 212555.156 121003 A 08 0 -0265
CCDCB603 01 24 00 01 00 01 00 01 CA7F0414 212555.156 121003 A 08 0 -0265
```
Fig. 5. DAQ output data example.

A typical data output example is shown in fig. 5. The first number in the first data line is the 32-bit CPLD count at the trigger time: C8B8E2A0 (hex) = 3367559840 (decimal). The next 8 hex numbers represent TDC edge timing information of the 4 input channels (one each for rising edge and falling edge). The 10th number is the 32-bit CPLD count of the most recent 1PPS leading edge: C8033BA6 (hex) = 3355655078 (decimal). The following numbers are time, date and GPS status info, extracted directly from the most recent GPS serial data in NMEA format: "212554.156" corresponds to UTC time 21:25:54 and 156 milliseconds, "121003" corresponds to October 12, 2003, and "A" means that the GPS status was "valid" (sufficient number of tracked satellites for position and time solution). The next field, "08", means 8 GPS satellites are tracked simultaneously, "0" means good DAQ status (no pending data or error bits), and finally, "-0266" represents the delay between the 1PPS leading edge and the start of the serial data set. In this case, the 1PPS pulse was ahead of the serial data by 266 ms.

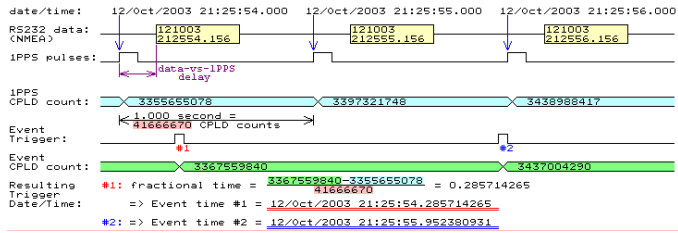

Fig. 6. Event time-stamping illustration.

Fig.6 is a timing diagram illustration of how the actual trigger time is computed, where event trigger pulse #1 corresponds to the first example line shown in fig.5. First, the CPLD oscillator frequency is computed by keeping track of the differences between the CPLD counts of consecutive 1PPS leading edges. Here, the average is 41666670 Hz. Secondly, the difference between the CPLD counts of the trigger leading edge and the 1PPS leading edge is computed and then divided by the averaged oscillator frequency. The result represents the fractional time duration of the trigger leading edge since the last 1PPS pulse; here: 0.285714265 seconds. Next, the UTC time down to integer seconds of the corresponding 1PPS pulse is computed; here: ROUND(212554.156 − 266/1000) = 212554, which is 21:25:54.000. Finally, these numbers are added to get the accurate UTC time of the event trigger #1: 21:25.54.285714265 on October 12, 2003 in the example.

## VII. DAQ PERFORMANCE

On October 11, 2003, two DAQ card sets were set up on the roof of the Physics-Astronomy building of the University of Washington to directly compare their time-stamping accuracy over a ~45-hour period. A test generator fed the inputs of both cards with identical short "event trigger" pulses at a rate of 1Hz. The GPS antennae of the two cards were placed on the roof such that each of them could only "see" a separate half of the sky and thus necessarily each used a different subset of GPS satellites for their location and time fixes.

Fig.7 shows the resulting time series plot of the time differences between the data of the two DAQ cards, and fig.8 is the corresponding histogram of the same data. The histogram shows a nearly gaussian distribution with a slight offset in the mean, which comes from systematic differences (mostly cable length) not yet compensated in the calculation. The standard deviation is 55 nanoseconds, and 99.9% of the data agrees within +/-350 ns between the two DAQ sets. This accuracy is more than sufficient for the school-network cosmic ray detector application where the DAQ cards are being used.

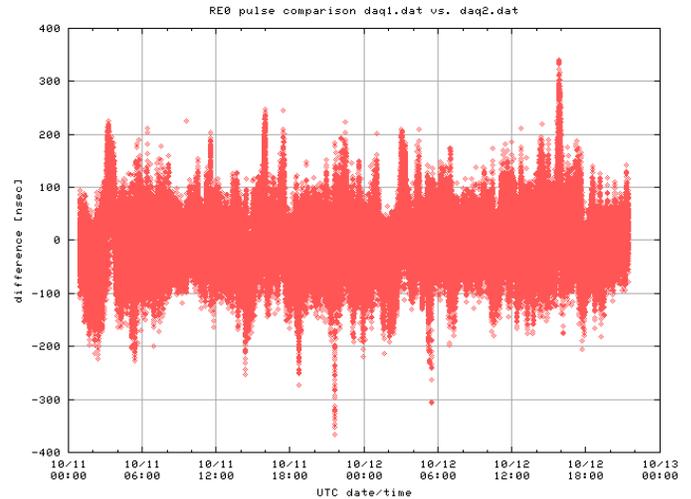

Fig. 7. Time series plot of the time-stamping comparison between two independent DAQ cards.

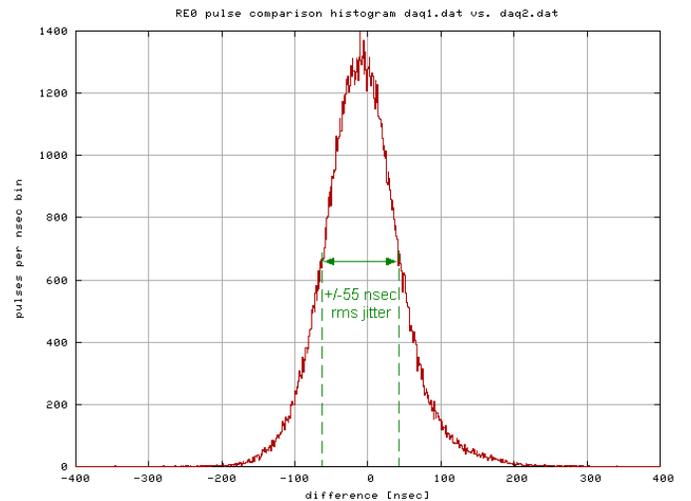

Fig. 8. Histogram of DAQ time-stamping comparison in Fig. 7.

## VIII. ACKNOWLEDGMENT

We wish to thank Tom Jordan, Sten Hansen and Terry Kiper at Fermilab, and Greg Snow and Dan Claes at the University of Nebraska (CROP) for their support and assistance.